\documentclass{ws-mpla}

\usepackage{graphicx,epsfig}
\usepackage{amsmath}
\usepackage{subfigure}
\usepackage{dcolumn}
\usepackage{amsmath}
\usepackage{amsfonts}
\usepackage{amssymb}
\usepackage{braket}
\usepackage{amssymb}
\usepackage{amsmath}
\usepackage{graphicx}
\usepackage{subfigure}
\usepackage{makeidx}
\usepackage{capt-of}
\usepackage{float}
\usepackage{placeins}
\usepackage{perpage}
\usepackage{mwe}
\usepackage{kantlipsum}
\usepackage{graphicx}
\usepackage{amsfonts}
\usepackage[colorlinks,linkcolor=blue,citecolor=red]{hyperref}

\begin{document}

%\markboth{Authors' Names}
%{Instructions for Typing Manuscripts (Paper's Title)}

%%%%%%%%%%%%%%%%%%%%% Publisher's Area please ignore %%%%%%%%%%%%%%%
%
%\catchline{}{}{}{}{}
%
%%%%%%%%%%%%%%%%%%%%%%%%%%%%%%%%%%%%%%%%%%%%%%%%%%%%%%%%%%%%%%%%%%%%

\title{On the origin of the holographic universe}

\author{Haidar Sheikhahmadi}

\address{School of Astronomy, Institute for Research in Fundamental Sciences (IPM),  P. O. Box 19395-5531, Tehran, Iran,\\
Center for Space Research, North-West University, Potchefstroom, South Africa,\\
Canadian Quantum Research Center 204-3002 32 Avenue Vernon, British Columbia V1T 2L7, Canada\\
h.sh.ahmadi@gmail.com~;~h.sheikhahmadi@ipm.ir}

\maketitle

\begin{history}
\received{(Day Month Year)}
\revised{(Day Month Year)}
\end{history}

\begin{abstract}
In this work, we reexamine the holographic dark energy concept proposed already for cosmological applications. By considering, more precisely, the bounds on the entropy arising from lattice field theory on one side and  Bekenstein-Hawking entropy of black holes on another side, it is shown that the so-called holographic dark energy cannot be mimicked as easily as claimed in the literature. In addition, the limits on the electron $(g-2)$ experiments are taken into account again. It is shown that the corrections to the electron magnetic momentum are of the order of ${\mathcal{O}}(10^{-23})$.\\
\textbf{keywords}:~{Holography concept; Lattice Field Theory; Bekenstein-Hawking entropy; Electron gyromagnetic anomaly}
\end{abstract}

%\tableofcontents

%% main text
%%%%%%%%%%%%%%%%%%%%%%%%%%%%%%%%%%%%%%%%
%%%%%%%%%%%%%%%%%%%%%%%%%%%%%%%%%%%%%
%%%%%%%%%%%%%%%%%%%%%%%%%%%%%%%%%%%%%%%%%%%%%
%%%%%%%%%%%%%%%%%%%%%%%%%%%%%%%%%%%%%%%%%%%%%
%%%%%%%%%%%%%%%%%%%%%%%%%%%%%%%%%%%%%%%%
%%%%%%%%%%%%%%%%%%%%%%%%%%%%%%%%%%%%%
%%%%%%%%%%%%%%%%%%%%%%%%%%%%%%%%%%%%%%%%%%%%%
\textcolor[rgb]{0.00,0.07,1.00}{\section{Introduction} \label{Introduction00}}
Determining at what scales both in length and energy, and consequently in mass, the correct interpretation of the concept of quantum gravity (at least still as a gedanken experiment) has been one of the most serious studies in the last half-century \cite{Isham:1993ji}. A model that is all-encompassing and can bring all the forces and interactions of nature under one umbrella, at least, is not yet built \cite{Garay:1994en}. In fact, the main problem goes back to the structure of the frame in which the remnant of the particles and the interactions are placed. Unlike flat space time, gravity, or in other words bend geometry, makes it possible for an active and consequently dynamic spacetime, that can interact with the rest of the contents of this framework. From the  point of view of quantum mechanics, any uncertainty in the position of the particle  implies an uncertainty in the momentum. In the presence of gravity and due to the gravity-energy interaction, we expect an additional uncertainty in the position of the particle. By means of a relatively straightforward and model-independent calculation, in the presence of gravity, it can be shown that the maximum distance between two events can be considered as small as the Plank length, i.e., $\sqrt{\bar{g} L}\leq L_{Pl},$ where we denote the gravitational potential by $\bar{g}$ and a wavelength of the photon  by $\simeq L$  where $L$ stands for the radius of a tesseract, for instance see \cite{Wheele:1964}. Now let us look at the  minimum length problem  by considering the vision arising from  loop quantum gravity, LQG. In a non-perturbative version of the LQG, we do not need to perturb the geometry itself, and therefore the background will not directly play a role in subsequent evolutions, but we can instead look for operators that carry information about the  structure of the geometry. In LQG, such quantum operators can be defined,  e.g., the surface operator, and are established in such a way that at lower energies, for example in effective field theory, EFT, limits they return to the classical geometry \cite{Rovelli:1989za,Ashtekar:1992tm}.The significance of such an interpretation can be understood in Bekenstein's work on Kerr black hole \cite{Bekenstein:1974jk}. This interpretation is important, in that we need a black hole entropy for our analysis,  that is proportional to the surface of the black hole, to constrain the infrared, IR, and ultraviolet, UV, cut-offs connection to eliminate the strong gravitational effects aiming at achieving an feasible EFT limit. As discussed in  \cite{Cohen:1998zx}, henceforth CKN, this IR-UV relationship  can supply a third possibility to overcome the cosmological constant problem. To interpret the consequences of such a connection, one can consider both the theoretical and experimental works in this regard, especially electron $(g-2)$ criterion \cite{Combley:1979ag,Morel:2020dww,Aoyama:2020ynm}. Basically,  the $(g-2)$   experiment allows us to calculate the deviations that occur due to the presence of gravity.  For a technical review on both the UV-IR cutoffs correlation and $(g-2)$ test, and to avoid prolonging the discussion, let us refer the reader to  \cite{Cohen:1998zx,Aoyama:2020ynm,Banks:2019arz,Cohen:2021zzr} and references therein.
Now we turn our attention to the so called holographic dark energy model obviously as an emergent  of CKN model. The concept of holography principle that raised from string theory, first was introduced in \cite{tHooft:1993dmi,tHooft:1999rgb} and \cite{Susskind:1994vu} aiming at solving shortcomings that appeared to construct  a feasible solution for quantum gravity. In such a proposal, it can be supposed that the description of a special volume of space on a lower-dimensional boundary to this region is achievable  \cite{Bousso:2002ju}. There are some outstanding examples of the applications of holography concept in high energy physics, of which the anti De Sitter/conformal field theory, AdS/CFT, correspondence \cite{Maldacena:1997re,Lin:2004nb} and entropy of black holes \cite{Bekenstein:1974jk,Bekenstein:1973ur}  are well-known to physics community.\\
{Here also it deserves to note that, the importance of the relation between mass  and UV cutoff from different perspectives, by taking into account for instance supersymmetry and supergravity, are addresses satisfactorily, where we refer the reader to \cite{Carroll:1998zi,Bellazzini:2019bzh,El-Nabulsi:2006bkk,Kawasaki:2011zi,El-Nabulsi:2005sss,Kallosh:2002gf} and references therein.}
The main motivation for doing this research is to re-examine the concept of holography, which has recently found a special place in the discussion of dark energy investigations. As will be seen, the proportion defined by the inverse of the area defined for the dark energy density will deviate.\\
This work is organised as follows:
Sec.\,\ref{Introduction00} was a brief introduction aiming to address the problems that the ill-defined holographic dark energy model is faced. In Sec.\, \ref{section1} the claimed problems are worked out then in order to constrain the cut-offs bounds  the $(g-2)$ criterion are utilized. Ultimately Sec.\, \ref{section2} is devoted to the discussion and conclusions of the work.\\

\textcolor[rgb]{0.00,0.07,1.00}{\section{Basic Properties of Holographic Universe}\label{section1}}
{With the help of the  quantum field theory, for a hypothetical box with the size of $L$ and the UV cutoff  $\Lambda$ the entropy will be an extensive quantity and reads $L^3\Lambda^3$ \cite{Cohen:1998zx}. This is the result that will be obtained with the help of the normalization in the lattice field theory as well.
On the other side, from the quantum states and their re-normalization  near the black hole horizon, from its standard point of view, the entropy of a black hole  through the relationship with the surface is closely related to  length of the black hole \cite{Bekenstein:1974jk,Bekenstein:1973ur}.  Now, based on the discussions appeared in the introduction, if we want to have an effective version of quantum field theory, say EFT, there must be a relationship between the entropy of the black hole and the state of the lattice field theory \cite{Cohen:1998zx,tHooft:1999rgb}. This actually guarantees that we will be at lower orders of magnitude that the Planck energy is, because as mentioned we do not have an achievable version of quantum gravity and we do need to work with effective theories. Besides these, such a relationship is necessary to guarantee the Bekenstein upper bounds aiming at avoiding the breakdown of black hole formation within the EFT context} based on \cite{Cohen:1998zx,Banks:2019arz,Cohen:2021zzr,tHooft:1999rgb,Guberina:2006qh,LopesCardoso:1998tkj}, there should be a bound for the entropy of black hole as
\begin{equation}\label{1a}
	L^3 \Lambda^3\lesssim (S_{BH})\equiv \pi M^{2}_{Pl} L^{2}\,,
\end{equation}
where  the  holographic Bekenstein-Hawking, HBH, entropy denoted by $S_{BH} \sim M^{2}_{Pl} L^2 $  and  Planck mass by $M_{Pl}$  \cite{Bekenstein:1973ur}.
In $L^3 \Lambda^3$ that is the lattice field theory, LFT, entropy, $L$ stands for  IR cutoff and $\Lambda$ for UV one, \textbf{and in large scale contexts stands for cosmological constant} \cite{Guberina:2006qh,LopesCardoso:1998tkj,Wei:2009kp,Sheikhahmadi:2014rka}.
To obtain a definition for energy density in large scale limits, one can multiply both sides of  Eq.\eqref{1a} by $\Lambda$ then readily it expresses
\begin{equation}\label{2a}
	L^3 \Lambda^4\lesssim \pi M^{2}_{Pl} L^{2}\Lambda\,,
\end{equation}
then by means of EFT techniques the energy density that corresponds to zero point quantum fluctuations, that appears  as $\Lambda^4$,  using \eqref{2a} can be rewritten as
\begin{equation}\label{3a}
	\rho_{\Lambda}=\Lambda^4 \lesssim \pi M_{Pl}^{2}L^{-1}\Lambda={3c^2 M_{Pl}^{2}}{L^{-1}}\Lambda\,,
\end{equation}
where even if IR cutoff behaves like $\Lambda^{-2}$, according to \cite{Cohen:1998zx,Cohen:2021zzr}, the holography concept can not be mimicked anymore, at least for a cosmological
purpose. The numerical constant $3c^2$  is introduced for convenience and it is not dimensionless as appeared usually in the literature. {Here also we should emphasis that, in some works this mater were declared that  the so called holographic parameter $c^2$ can not be considered as a constant in general, see  \cite{Radicella:2010vf,El-Nabulsi:2009cdp,Sheykhi:2016emj}}.
Comparing Eq.\eqref{3a} to its analogues that appeared in  \cite{Cohen:1998zx} and \cite{Cohen:2021zzr}, it is not allowed to easily consider this equation as a holographic dark energy equation,  that should mimic such a formula $\Lambda^4=\rho_\Lambda \approxeq \pi M^{2}_{Pl} L^{-2}$. In this regard, as discussed in CKN and CK, \cite{Cohen:1998zx,Cohen:2021zzr}, and obviously from  Eq.\eqref{1a}, the IR and UV cutoffs can not be treated as independent parameters. Therefore, one cannot introduce an arbitrary scale for these cutoffs aiming at getting rid of this  shortcoming, appeared due to inequality \eqref{2a} here. In this regard, to reconstruct holographic dark energy model, one maybe would like to rescale the length as $L\propto \Lambda^{-1}$, but obviously it is forbidden based on aforementioned discussions, inequality \eqref{2a} and the following explanations. These results  besides Eq.\eqref{3a} and \eqref{4a} are important because they set a boundary on which the EFT rules can be employed  properly.  In other words, they determine how and where the effects of quantum gravity can be relinquished. \\
Now let's turn our attention to the so called electron $(g-2)$ experiment. Following CKN and CK, \cite{Cohen:1998zx,Cohen:2021zzr}, and by utilizing Eq.\eqref{1a} we can obtain the constraint between IR and UV cutoffs as follows,
\begin{equation}\label{4a}
	L \Lambda^3\lesssim \pi M^{2}_{Pl}\,.
\end{equation}
This relation indicates a correlation between the energy and the size of the box, that is the Universe, and such a coupling remains for even perturbative EFT calculations. The analogues of this equation  appeared in CK paper as
\begin{equation}\label{4b}
	L \Lambda^2\lesssim M_{Pl}\,,
\end{equation}
and obviously Eqs.\eqref{4a} and \eqref{4b} behave completely different, as it is shown, in determining the amount of  bonds on the energy and the size of the box. In fact, the problem goes back to ignoring $L \Lambda$, by CKN and CK, on the right hand side of Eq.\eqref{2a}, which allows getting a square root of both sides of this equation, and therefore allowed them to claim a holographic like dark energy model for the Universe.\\
{As discussed, the relationship between the  UV and IR cutoffs and how they are renormalized play an important role in investigating the thermodynamics of a black hole and especially its entropy \cite{Davies:1976pm,Bunch:1978yq}. In fact, many works have been done to investigate the vacuum polarization around a black hole, all of which were to obtain a finite value for the expectation value of the energy density and its relationship with the temperature and finally the entropy of the black hole \cite{Christensen:1976vb,Candelas:1980zt,Candelas:1984pg}. On the other hand, in the asymptotic limits and far from the horizon of the black hole, investigations are also important for the role of the IR cutoff and its renormalization, which may even include interesting phenomena such as the cosmological constant \cite{Firouzjahi:2022vij}.}

In the next stage, we want to examine this model in the concordance the well known electron gyromagnetic anomaly terrestrial experiment.\\
\textcolor[rgb]{0.00,0.00,1.00}{\subsection{Constraints from g-2 experiment}}
Here, following CKN and CK, we want to calculate the corrections to the usual calculations by virtue of the $(g-2)$ experiment. In order to reexamine such calculations one can consider the $1$-loop results for electron
gyromagnetic anomaly, $a=(g-2)/2$,  for a lepton like electron of mass $m$, i.e.
\begin{equation}\label{4c}
	a(L, \Lambda)=\left(\frac{\alpha}{2 \pi}\right)\left(1-\frac{\pi^{2}}{m L}-\frac{m^{2}}{3 \Lambda^{2}}+\cdots\right)\,,
\end{equation}
see \cite{Cohen:1998zx,Cohen:2021zzr}. As discussed by CK,  to eliminate the effects of strong gravitational effects and consequently the sensitivity of the model, to such gravitational effects, one possible way is minimizing IR-UV corrections.  In doing so by utilizing Eqs.\eqref{4a} and \eqref{4c} it leads to the following results for $\Lambda$ and $L$,
\begin{equation}\label{5a}
	\Lambda\sim m_{eff}^{2/3}\left(\frac{M_{Pl}^2}{m}\right)^{\frac{1}{6}}~{GeV}\,,
\end{equation}
and
\begin{equation}\label{5b}
	L\sim \pi m^{-1}\left(\frac{M_{Pl}^2}{\tilde{m}_{eff}}\right)^{\frac{1}{2}}~{GeV^{-1}}\,,
\end{equation}
where $m_{eff}\equiv\kappa  m$ and $\tilde{m}_{eff}\equiv \lambda m$ are introduced for dimensional reasons and $\kappa$ and $\lambda$ are constants of the order of unity. Comparing the results  in Eqs.\eqref{5a} and \eqref{5b} and their analogues in CKN or CK, one immediately observe that they are not as same as each other.
By introducing dimension of $[\kappa]\propto\mathcal{O}(MeV^{1/4})$ and $[\lambda]\propto\mathcal{O}(MeV)$, and considering $m$ as electron mass, the amount of $\Lambda$ and $L$  immediately read
\begin{equation}\label{5c}
	\Lambda\simeq{10~GeV}\,, \,\,\,\,\,\,\,\,\,\,\, L\simeq{200~cm}\,.
\end{equation}
In CK, they obtained respectively $\Lambda\simeq{200~GeV}$ and $L\simeq{6~cm}$.
Then from Eqs.\eqref{4c},\eqref{5a} and \eqref{5b} for the minimum deviation from the Schwinger $1-$loop result one gets
\begin{equation}\label{6a}
	a-\frac{\alpha }{2\pi} \sim \frac{\alpha}{2\pi}\left(\pi\sqrt{\frac{\tilde{m}_{eff}}{M_{Pl}^2}}+\frac{m^{2/3}}{3\kappa^{4/3}}\sqrt[3]{\frac{m}{M_{Pl}^2}}\right)\sim\mathcal{O}(10^{-23})\,,
\end{equation}
that is different comparing to the results of \cite{Cohen:1998zx,Cohen:2021zzr}.
\\
\textcolor[rgb]{0.00,0.07,1.00}{\section{Concluding remarks}\label{section2}}
Our main aim of doing this study was to check the validity of the proposed holographic dark energy model that appeared in the literature and especially in CKN paper.  It has been shown that, by considering precisely the bounds on the entropies arising from LFT and black holes the so-called holographic dark energy model can not be claimed anymore. It has been also figured out that, it is not arbitrary, to re-scale the UV cutoff nor like $\propto L^{-2}$ neither $\propto L^{-1}$  as discussed in CKN, CK or references therein. The consequences of this correction obviously affect the so-called entropy-corrected holographic dark energy \cite{Wei:2009kp}, agegraphic dark energy \cite{Cai:2007us}, new agegraphic dark energy \cite{Wei:2007ut} and other emergent prototypes.  This new constraint which has appeared in \eqref{4a} is examined to check the order of the theoretical corrections that appeared in the electron $(g-2)$ experiment and it was of the order of ${\mathcal{O}}(10^{-23})$. Ultimately, in this report, a correct version of the UV-IR bound resulted in a new version for holographic-like dark energy models obtained.\\
\emph{Examining the corrections that will appear for the aforementioned emergent  models through this study will be the subject of future work.}
\\
\\
\textcolor[rgb]{0.00,0.00,1.00}{\section*{Acknowledgments}}
H.S. is very grateful to the anonymous referees for their valuable comments resulted in improvement of the paper.   H.S. would like also to thank H. Firouzjahi  for very constructive and interesting discussions  on the vacuum polarization around black holes and its renormalization. He is grateful to T. Harko for the discussions  on the original draft of the work.
\\


\begin{thebibliography}{}
	
	
	
	
	
	
	%\cite{Isham:1993ji}
	\bibitem{Isham:1993ji}
	C.~J.~Isham,
	``Prima facie questions in quantum gravity,''
	Lect. Notes Phys. \textbf{434}, 1-21 (1994)
	doi:10.1007/3-540-58339-4\_13
	[arXiv:gr-qc/9310031 [gr-qc]].
	%57 citations counted in INSPIRE as of 28 Sep 2021
	
	%\cite{Garay:1994en}
	\bibitem{Garay:1994en}
	L.~J.~Garay,
	``Quantum gravity and minimum length,''
	Int. J. Mod. Phys. A \textbf{10}, 145-166 (1995)
	doi:10.1142/S0217751X95000085
	[arXiv:gr-qc/9403008 [gr-qc]].
	%1143 citations counted in INSPIRE as of 28 Sep 2021
	
	\bibitem{Wheele:1964} J. A. Wheeler, in Relativity, Groups and Topology, eds. B. S. DeWitt and C. DeWitt
	(Gordon and Breach, London, 1964). Geometrodynamics and the issue of the final state.
	
	%\cite{Rovelli:1989za}
	\bibitem{Rovelli:1989za}
	C.~Rovelli and L.~Smolin,
	``Loop Space Representation of Quantum General Relativity,''
	Nucl. Phys. B \textbf{331}, 80-152 (1990)
	doi:10.1016/0550-3213(90)90019-A
	%713 citations counted in INSPIRE as of 28 Sep 2021
	
	%\cite{Ashtekar:1992tm}
	\bibitem{Ashtekar:1992tm}
	A.~Ashtekar, C.~Rovelli and L.~Smolin,
	``Weaving a classical geometry with quantum threads,''
	Phys. Rev. Lett. \textbf{69}, 237-240 (1992)
	doi:10.1103/PhysRevLett.69.237
	[arXiv:hep-th/9203079 [hep-th]].
	%330 citations counted in INSPIRE as of 28 Sep 2021
	
	
	%\cite{Bekenstein:1974jk}
	\bibitem{Bekenstein:1974jk}
	J.~D.~Bekenstein,
	``The quantum mass spectrum of the Kerr black hole,''
	Lett. Nuovo Cim. \textbf{11}, 467 (1974)
	doi:10.1007/BF02762768
	%461 citations counted in INSPIRE as of 28 Sep 2021
	
	
	
	%%%%%%%%%%%%%% g-2
	
	
	%\cite{Cohen:1998zx}
	\bibitem{Cohen:1998zx}
	A.~G.~Cohen, D.~B.~Kaplan and A.~E.~Nelson,
	``Effective field theory, black holes, and the cosmological constant,''
	Phys. Rev. Lett. \textbf{82}, 4971-4974 (1999)
	doi:10.1103/PhysRevLett.82.4971
	[arXiv:hep-th/9803132 [hep-th]].
	%1050 citations counted in INSPIRE as of 09 Sep 2021
	
	
	
	
	
	
	\bibitem{Combley:1979ag} F. H.Combley,
	"(g-2) factors for muon and electron and the consequences for QED",
	Reports on Progress in Physics.\textbf{ 42 }(12), 1889 (1979).
	
	%\cite{Morel:2020dww}
	\bibitem{Morel:2020dww}
	L.~Morel, Z.~Yao, P.~Clad\'e and S.~Guellati-Kh\'elifa,
	``Determination of the fine-structure constant with an accuracy of 81 parts per trillion,''
	Nature \textbf{588}, no.7836, 61-65 (2020)
	doi:10.1038/s41586-020-2964-7
	%83 citations counted in INSPIRE as of 29 Sep 2021
	
	
	%\cite{Aoyama:2020ynm}
	\bibitem{Aoyama:2020ynm}
	T.~Aoyama, N.~Asmussen, M.~Benayoun, J.~Bijnens, T.~Blum, M.~Bruno, I.~Caprini, C.~M.~Carloni Calame, M.~C\`e and G.~Colangelo, \textit{et al.}
	``The anomalous magnetic moment of the muon in the Standard Model,''
	Phys. Rept. \textbf{887}, 1-166 (2020)
	doi:10.1016/j.physrep.2020.07.006
	[arXiv:2006.04822 [hep-ph]].
	%351 citations counted in INSPIRE as of 29 Sep 2021
	
	
	
	
	%\cite{Banks:2019arz}
	\bibitem{Banks:2019arz}
	T.~Banks and P.~Draper,
	``Remarks on the Cohen-Kaplan-Nelson bound,''
	Phys. Rev. D \textbf{101}, no.12, 126010 (2020)
	doi:10.1103/PhysRevD.101.126010
	[arXiv:1911.05778 [hep-th]].
	%5 citations counted in INSPIRE as of 13 Sep 2021
	
	
	%\cite{Cohen:2021zzr}
	\bibitem{Cohen:2021zzr}
	A.~G.~Cohen and D.~B.~Kaplan,
	``Gravitational contributions to the electron $g$-factor,''
	[arXiv:2103.04509 [hep-ph]].
	%5 citations counted in INSPIRE as of 09 Sep 2021
	
	%%%%%%%%%%%%%%%%%Holography Principle
	
	
	
	%\cite{tHooft:1993dmi}
	\bibitem{tHooft:1993dmi}
	G.~'t Hooft,
	``Dimensional reduction in quantum gravity,''
	Conf. Proc. C \textbf{930308}, 284-296 (1993)
	[arXiv:gr-qc/9310026 [gr-qc]].
	%2580 citations counted in INSPIRE as of 20 Sep 2021
	
	
	
	%\cite{tHooft:1999rgb}
	\bibitem{tHooft:1999rgb}
	G.~'t Hooft,
	``The Holographic principle: Opening lecture,''
	Subnucl. Ser. \textbf{37}, 72-100 (2001)
	doi:10.1142/9789812811585\_0005
	[arXiv:hep-th/0003004 [hep-th]].
	%146 citations counted in INSPIRE as of 09 Sep 2021
	
	
	
	
	
	
	%\cite{Susskind:1994vu}
	\bibitem{Susskind:1994vu}
	L.~Susskind,
	``The World as a hologram,''
	J. Math. Phys. \textbf{36}, 6377-6396 (1995)
	doi:10.1063/1.531249
	[arXiv:hep-th/9409089 [hep-th]].
	%3153 citations counted in INSPIRE as of 09 Sep 2021
	
	
	
	
	
	
	
	
	%\cite{Bousso:2002ju}
	\bibitem{Bousso:2002ju}
	R.~Bousso,
	``The Holographic principle,''
	Rev. Mod. Phys. \textbf{74}, 825-874 (2002)
	doi:10.1103/RevModPhys.74.825
	[arXiv:hep-th/0203101 [hep-th]].
	%1113 citations counted in INSPIRE as of 09 Sep 2021
	
	
	%%%%%%%%%%%%%%%%%%%%%%AdS/CFT
	
	
	%\cite{Maldacena:1997re}
	\bibitem{Maldacena:1997re}
	J.~M.~Maldacena,
	``The Large N limit of superconformal field theories and supergravity,''
	Adv. Theor. Math. Phys. \textbf{2}, 231-252 (1998)
	doi:10.1023/A:1026654312961
	[arXiv:hep-th/9711200 [hep-th]].
	%16921 citations counted in INSPIRE as of 09 Sep 2021
	
	
	%\cite{Lin:2004nb}
	\bibitem{Lin:2004nb}
	H.~Lin, O.~Lunin and J.~M.~Maldacena,
	``Bubbling AdS space and 1/2 BPS geometries,''
	JHEP \textbf{10}, 025 (2004)
	doi:10.1088/1126-6708/2004/10/025
	[arXiv:hep-th/0409174 [hep-th]].
	%778 citations counted in INSPIRE as of 09 Sep 2021
	
	
	
	%%%%%%%%%%%%%%%%%Entropy of Black holes
	
	
	%\cite{Bekenstein:1973ur}
	\bibitem{Bekenstein:1973ur}
	J.~D.~Bekenstein,
	``Black holes and entropy,''
	Phys. Rev. D \textbf{7}, 2333-2346 (1973)
	doi:10.1103/PhysRevD.7.2333
	%5221 citations counted in INSPIRE as of 09 Sep 2021
	
	
	%%%%%%%%%%%%%%%%Entropy Corrected HDE
	
	
	
%\cite{Carroll:1998zi}
\bibitem{Carroll:1998zi}
S.~M.~Carroll,
``Quintessence and the rest of the world,''
Phys. Rev. Lett. \textbf{81}, 3067-3070 (1998)
doi:10.1103/PhysRevLett.81.3067
[arXiv:astro-ph/9806099 [astro-ph]].
%1053 citations counted in INSPIRE as of 09 Jan 2023

%\cite{Bellazzini:2019bzh}
\bibitem{Bellazzini:2019bzh}
B.~Bellazzini, F.~Riva, J.~Serra and F.~Sgarlata,
``Massive Higher Spins: Effective Theory and Consistency,''
JHEP \textbf{10}, 189 (2019)
doi:10.1007/JHEP10(2019)189
[arXiv:1903.08664 [hep-th]].
%37 citations counted in INSPIRE as of 09 Jan 2023

%\cite{El-Nabulsi:2006bkk}
\bibitem{El-Nabulsi:2006bkk}
R.~A.~El-Nabulsi,
``Phase transitions in the early universe with negatively induced supergravity cosmological constant,''
Chin. Phys. Lett. \textbf{23}, 1124-1127 (2006)
doi:10.1088/0256-307X/23/5/017
%19 citations counted in INSPIRE as of 09 Jan 2023

%\cite{Kawasaki:2011zi}
\bibitem{Kawasaki:2011zi}
M.~Kawasaki and T.~Takesako,
``Hubble Induced Mass in Radiation Dominated Universe,''
Phys. Lett. B \textbf{711}, 173-177 (2012)
doi:10.1016/j.physletb.2012.03.069
[arXiv:1112.5823 [hep-ph]].
%16 citations counted in INSPIRE as of 09 Jan 2023

%\cite{El-Nabulsi:2005sss}
\bibitem{El-Nabulsi:2005sss}
R.~A.~El-Nabulsi,
``Effective cosmological constant from supergravity arguments and non-minimal coupling,''
Phys. Lett. B \textbf{619}, 26-29 (2005)
doi:10.1016/j.physletb.2005.06.002
%28 citations counted in INSPIRE as of 09 Jan 2023

%\cite{Kallosh:2002gf}
\bibitem{Kallosh:2002gf}
R.~Kallosh, A.~D.~Linde, S.~Prokushkin and M.~Shmakova,
``Supergravity, dark energy and the fate of the universe,''
Phys. Rev. D \textbf{66}, 123503 (2002)
doi:10.1103/PhysRevD.66.123503
[arXiv:hep-th/0208156 [hep-th]].
%106 citations counted in INSPIRE as of 09 Jan 2023

	
	
	
	
	%\cite{Guberina:2006qh}
	\bibitem{Guberina:2006qh}
	B.~Guberina, R.~Horvat and H.~Nikolic,
	``Nonsaturated Holographic Dark Energy,''
	JCAP \textbf{01}, 012 (2007)
	doi:10.1088/1475-7516/2007/01/012
	[arXiv:astro-ph/0611299 [astro-ph]].
	%90 citations counted in INSPIRE as of 09 Sep 2021
	
	
	
	%\cite{LopesCardoso:1998tkj}
	\bibitem{LopesCardoso:1998tkj}
	G.~Lopes Cardoso, B.~de Wit and T.~Mohaupt,
	``Corrections to macroscopic supersymmetric black hole entropy,''
	Phys. Lett. B \textbf{451}, 309-316 (1999)
	doi:10.1016/S0370-2693(99)00227-0
	[arXiv:hep-th/9812082 [hep-th]].
	%286 citations counted in INSPIRE as of 09 Sep 2021
	
	
	%\cite{Wei:2009kp}
	\bibitem{Wei:2009kp}
	H.~Wei,
	``Entropy-Corrected Holographic Dark Energy,''
	Commun. Theor. Phys. \textbf{52}, 743-749 (2009)
	doi:10.1088/0253-6102/52/4/35
	[arXiv:0902.0129 [gr-qc]].
	%112 citations counted in INSPIRE as of 10 Sep 2021
	
	%\cite{Sheikhahmadi:2014rka}
	\bibitem{Sheikhahmadi:2014rka}
	H.~Sheikhahmadi, A.~Aghamohammadi and K.~Saaidi,
	``The effect of de Sitter like background on increasing the zero point budget of dark energy,''
	Adv. High Energy Phys. \textbf{2016}, 2594189 (2016)
	doi:10.1155/2016/2594189
	[arXiv:1407.0125 [gr-qc]].
	%5 citations counted in INSPIRE as of 09 Sep 2021
	
	
	



%\cite{Radicella:2010vf}
\bibitem{Radicella:2010vf}
N.~Radicella and D.~Pavon,
``On the $c^{2}$ term in the holographic formula for dark energy,''
JCAP \textbf{10}, 005 (2010)
doi:10.1088/1475-7516/2010/10/005
[arXiv:1007.4129 [gr-qc]].
%43 citations counted in INSPIRE as of 09 Jan 2023




%\cite{El-Nabulsi:2009cdp}
\bibitem{El-Nabulsi:2009cdp}
R.~A.~El-Nabulsi,
``Maxwell brane cosmology with higher-order string curvature corrections, a nonminimally coupled scalar field, dark matter-dark energy interaction and a varying speed of light,''
Int. J. Mod. Phys. D \textbf{18}, 289-318 (2009)
doi:10.1142/S0218271809014431
%16 citations counted in INSPIRE as of 09 Jan 2023

%\cite{Sheykhi:2016emj}
\bibitem{Sheykhi:2016emj}
A.~Sheykhi, S.~Ghaffari and N.~Roshanshah,
``A note on holographic dark energy with varying $c^2$ term,''
Int. J. Theor. Phys. \textbf{56}, no.6, 1845-1860 (2017)
doi:10.1007/s10773-017-3329-3
[arXiv:1612.03040 [physics.gen-ph]].
%2 citations counted in INSPIRE as of 09 Jan 2023


%\cite{Davies:1976pm}
\bibitem{Davies:1976pm}
P.~C.~W.~Davies,
``Quantum Field Theory in Curved Space-Time,''
Nature \textbf{263}, 377-380 (1976)
doi:10.1038/263377a0
%10 citations counted in INSPIRE as of 09 Jan 2023
%\cite{Bunch:1978yq}
\bibitem{Bunch:1978yq}
T.~S.~Bunch and P.~C.~W.~Davies,
``Quantum Field Theory in de Sitter Space: Renormalization by Point Splitting,''
Proc. Roy. Soc. Lond. A \textbf{360}, 117-134 (1978)
doi:10.1098/rspa.1978.0060
%1150 citations counted in INSPIRE as of 09 Jan 2023
	


%\cite{Christensen:1976vb}
\bibitem{Christensen:1976vb}
S.~M.~Christensen,
``Vacuum Expectation Value of the Stress Tensor in an Arbitrary Curved Background: The Covariant Point Separation Method,''
Phys. Rev. D \textbf{14}, 2490-2501 (1976)
doi:10.1103/PhysRevD.14.2490
%510 citations counted in INSPIRE as of 11 Jan 2023	


%\cite{Candelas:1980zt}
\bibitem{Candelas:1980zt}
P.~Candelas,
``Vacuum Polarization in Schwarzschild Space-Time,''
Phys. Rev. D \textbf{21}, 2185-2202 (1980)
doi:10.1103/PhysRevD.21.2185
%363 citations counted in INSPIRE as of 11 Jan 2023
	
	

%\cite{Candelas:1984pg}
\bibitem{Candelas:1984pg}
P.~Candelas and K.~W.~Howard,
``VACUUM (PHI**2) IN SCHWARZSCHILD SPACE-TIME,''
Phys. Rev. D \textbf{29}, 1618-1625 (1984)
doi:10.1103/PhysRevD.29.1618
%88 citations counted in INSPIRE as of 11 Jan 2023



	

%\cite{Firouzjahi:2022vij}
\bibitem{Firouzjahi:2022vij}
H.~Firouzjahi,
``Cosmological constant and vacuum zero point energy in black hole backgrounds,''
Phys. Rev. D \textbf{106}, no.4, 045015 (2022)
doi:10.1103/PhysRevD.106.045015
[arXiv:2205.06561 [gr-qc]].
%2 citations counted in INSPIRE as of 09 Jan 2023


%%%%%%%%%%%%%%%%%%%%%%%%%%%%%%%%%%%%%%ADE & NADE  Cai:2007us,Wei:2007ut
	%\cite{Cai:2007us}
	\bibitem{Cai:2007us}
	R.~G.~Cai,
	``A Dark Energy Model Characterized by the Age of the Universe,''
	Phys. Lett. B \textbf{657}, 228-231 (2007)
	doi:10.1016/j.physletb.2007.09.061
	[arXiv:0707.4049 [hep-th]].
	%414 citations counted in INSPIRE as of 08 Sep 2021
	
	
	
	
	%\cite{Wei:2007ut}
	\bibitem{Wei:2007ut}
	H.~Wei and R.~G.~Cai,
	``Interacting Agegraphic Dark Energy,''
	Eur. Phys. J. C \textbf{59}, 99-105 (2009)
	doi:10.1140/epjc/s10052-008-0799-8
	[arXiv:0707.4052 [hep-th]].
	%152 citations counted in INSPIRE as of 08 Sep 2021
	
	
	%%%%%%%%%%%%%%%%%%%%%%%%%%%%%%%%%%%%%%%%%%%%%%%%%%%%%%%%%%%%
	
\end{thebibliography}
\end{document}